# Enhancing Resilience: Model-based Simulations[1]

**d'Artis Kancs**
Science Research Innovation Implementation Centre, Latvian National Armed Forces

d'artis.kancs@ec.europa.eu

*ABSTRACT*

*Since several years, the fragility of global supply chains (GSCs) is at historically high levels. In the same time, the landscape of hybrid threats is expanding; new forms of hybrid threats create different types of uncertainties. This paper aims to understand the potential consequences of uncertain events – like natural disasters, pandemics, hybrid and/or military aggression – on GSC resilience and robustness. Leveraging a parsimonious supply chain model, we analyse how the organisational structure of GSCs interacts with uncertainty, and how risk-aversion vs. ambiguity-aversion, vertical integration vs. upstream outsourcing, resilience vs. efficiency trade-offs drive a wedge between decentralised and centralised optimal GSC diversification strategies in presence of externalities. Parameterising the scalable data model with World-Input Output Tables, we simulate the survival probability of a GSC and implications for supply chain robustness and resilience. The presented model-based simulations provide an interoperable and directly comparable conceptualisation of positive and normative effects of counterfactual resilience and robustness policy choices under individually optimal (decentralised) and socially optimal (centralised) GSC organisation structures.*

## 1.0   INTRODUCTION

Two developments with a global character and dynamically interrelated across industries and countries have accelerated in recent years. One is an increasing vulnerability of global production networks. In the same time, the landscape of hybrid threats is expanding and intensifying (European Commission 2021). More than ever since the end of the Cold War, the last authoritarian regimes and strategic competitors test the Alliance's resilience and seek to exploit the openness, interconnectedness and digitalisation of free and open societies, interfere in democratic processes and institutions, and target the security of citizens through hybrid tactics, such as, the recent attempts of energy weaponisation by Russia. In the age of globalisation and cross-border production networks, lower cost is usually associated with economies of scale in input sourcing, whereas higher quality inputs tend to be found in markets with niche expertise - both implying higher participation in global supply chains (GSCs). Global production fragmentation increases foreign exposure of domestic industries, which participate extensively in GSCs. The specialisation and cost advantages for international companies that arise from involvement in GSCs are unavoidably associated with greater risks and ambiguity in the face of shocks, such as global pandemics, the climate crisis and hybrid attacks. These uncertainties are acknowledged by the Secretary General Stoltenberg: "over-reliance on the import of key commodities, like energy [on the sourcing-side, and] exporting advanced technologies, like Artificial Intelligence [on the selling-side] can create vulnerabilities and weakened resilience".

---

[1] The authors acknowledge helpful comments from Etienne Vincent as well as participants of the NATO Computer Assisted Analysis, Exercise, Experimentation (CA2X2) Forum in Rome organised by the NATO Modelling and Simulation Centre of Excellence; NATO Innovation Challenge finalists presentations in Bucharest organised by the NATO Innovation Hub  the NATO Supreme Allied Commander Transformation; NATO Operations Research and Analysis (OR&A) Conference in Copenhagen organised by the NATO Supreme Allied Commander Transformation and NATO Science and Technology Organization; NATO Emerging and Disruptive Modelling and Simulation Technologies Symposium in Bath organised by Modelling and Simulation Group, NATO Science and Technology Organization; International Research Conference GlobState 2022 in Bydgoszcz organised by the Doctrine and Training Centre of the Polish Armed Forces. The authors are solely responsible for the content of the paper. The views expressed are purely those of the authors and may not in any circumstances be regarded as stating an official position of the NATO, European Commission or the Latvian National Armed Forces. Any remaining errors are solely ours.

The first line of the Alliance's defence is resilience – ensuring that the socio-politico-economic fabric can function in the face of adversity.[2] Leveraging the strong commitment to action and achieving the desired capacity of resilience and robustness requires a holistic, integrated and dynamically coordinated approach. On the policy side, political leaders have to take the responsibility for being fully open with citizens about the changing character of hybrid threats. Achieving a socio-politico-economic resilience that meets the seven baseline requirements – which must be maintained under the most demanding circumstances – will require a mobilisation of resources. A full transparency is therefore paramount regarding the costs and sacrifices that will be needed, for example, to defend security in the face of Russia's war on Democracy and possible future warfare. As noted by Secretary General Stoltenberg at the World Economic Forum 2022: "we should not trade long-term security needs for short-term economic interests",[3] which implies costs and sacrifices.

How to 'achieve the required resilience' while doing as little damage as possible to the society's socio-politico-economic fabric'? Indeed, the challenge is to achieve long-term security goals without neglecting the short- and medium-term economic needs of economy and society. In the context of GSCs, the challenge is to ensure resilient and diversified supply chains in place to allow for a continued flow of essential goods and avoid shortages in the short-, medium- and long-run. Our analysis investigates this trade-off formally by framing it as a constrained optimisation problem with two constraints – a resilience/robustness constraint on the desirability side and a resource mobilisation constraint on the feasibility side. We simulate the optimal strategy of private sector firms in presence of GSC shocks under uncertainty. By integrating predictive analytics, model-based simulations provide interoperable and directly comparable quantifications of positive and normative effects of counterfactual resilience and robustness policy choices in critical and non-critical sectors. The scalable data model allows to identify strategies for addressing Alliance's vulnerabilities arising from societies' openness and economies' interconnectedness in international trade and global production networks, and the embedded information awareness tool facilitates strategic decision making.

The present study builds on and complements the existing Science & Technology Organisation (STO) strategic analytical support, including the Multi-Dimensional Data Farming, Causal Reasoning, and resilience tools. The Resilience Data Analytics Tool can be used, among others, to assess the levels of resilience by leveraging open-source data, big data analytics, machine learning, and data visualisation and allows the identification of potential shocks to the Alliance's resilience. The Resilience Model provides a holistic framework for simulating a wide range of Political, Military, Economic, Social, Information, and Infrastructure (PMESII) shocks (e.g. electricity blackout, cyber attack, martial law enforcement, big human movement, state of war, armed conflict), and allows assessing both resilience domains (civil support to the military, continuity of government, and continuity of essential services) as well as risk (command and control, protection, movement/ manoeuvre, and sustain) (Hodicky et al. 2020). The Joint Warfare Centre (JWC) leverages the Joint Theatre Level Simulation (JTLS). Our scalable data model - which is based on Antras and de Gortari (2020) and Jiang et al. (2022) - is complementary to the existing resilience modelling and simulation tools, as it is specifically designed to account for the asymmetric exposure of specific sectors, and to study the allies' resilience and robustness in the presence of exogenous shocks under uncertainty causing, for example, supply ruptures, demand ruptures/surges or transportation ruptures.

## 2.0 GLOBAL SUPPLY CHAINS: EVIDENCE

We begin by taking stocks of the Alliance's foreign input dependence. Knowing economies' international exposure on intermediates input and output markets is crucial to identify potential vulnerabilities and allow for policy actions where needs arise before major GSC-disruptions occur. We approach this question by querying recent statistical data through different micro-macro lenses.

---

[2] www.nato.int/docu/review/articles/2019/02/27/resilience-the-first-line-of-defence/

[3] www.nato.int/cps/en/natohq/opinions_195755.htm

First, we look at the firm-level perspective by using firm-level data to understand firms' input sourcing and output market decisions, how import and export participations are linked, and how globally operating firms organise their production networks. Second, a value-added (macro) approach is employed to assess how the industrial production is allocated internationally and how each stage of production contributes to the final product. This aggregated approach provides an alternative to the micro view by concentrating on countries and industries as the unit of analysis. The macro-approach to measuring GSCs connects national Input Output (IO) tables across borders using bilateral trade data to construct a World Input-Output Tables. These data are applied to measure trade in value added, as well as the length and location of producers in GSCs. Both in the firm-level analysis and value-added approach an understanding of the domestic economy's foreign exposure via the GSC channel requires a knowledge of where are goods made? This core question is approached from different sides (location of output, inputs) and by viewing through different lenses (micro, macro).

## 2.1 Micro perspective: Firm-level foreign reliance

We start with the micro approach, where the firms are the unit of analysis; they are the ones that decide whether and to what extent to participate in GSCs. Firms upstream and downstream face contracting problems – moral hazard or incomplete contracts. Integrating internationally and vertically helps to solve the informational problem and reduce supply chain uncertainty. Firm-level forward participation in GSCs is evidenced through exporting of intermediate inputs whereas backward participation in GSCs through importing of intermediate inputs. The observed world trade flows in international trade statistics are interpreted as the aggregation of individual firm-level decisions related to the destinations to which firms export their intermediate goods, but also the origin countries from which they source intermediate inputs, or the 'platform' countries from which they assemble goods for distant destination countries.

How exposed are Alliance's companies to foreign input supplies and output markets? This question can be answered at several levels. Taking the airspace industry as an example, we can say it was made in Stevenage when a product rolls off the Airbus Defence and Space assembly line in Stevenage, UK. This is the first-level truth, but it is not the whole truth. The second level recognises that the Stevenage plant buys inputs from other sectors located at home and abroad. Tracing the first-level production location of inputs gives us the second-level answer; this provides a directly observable dependence on foreign inputs. The intermediate import measure is directly observable in standard trade databases, and it has a number of advantages. However, this measure of intermediate imports is not the whole truth either because purchased foreign inputs also use inputs. The third-level answer - the whole truth of foreign input reliance - takes account of the entire recursive sequence of all the inputs into all the inputs. According to Lund et al. (2020), Airbus has 1,676 publicly disclosed tier-one suppliers. In the same time, Airbus works with over 12,000 tier-two suppliers and below worldwide. This implies that Airbus has more than 8 times as many total suppliers as in tier one.

An example of GSC interdependencies in the IT industry are illustrated in Figure 1, where tier-one and tier-two input suppliers of Dell [Military and Defence] (left) and Lenovo (right) are depicted. According to the Bloomberg Global Supply Chain Data, Dell has nearly 5,000 worldwide tier-one and tier-two input suppliers (second-level input interdependencies), whereas Lenovo draws on around 4,000 tier-one and tier-two input suppliers globally. More importantly, there are 2,272 shared input suppliers between these two US Chinese IT companies. These examples illuminate: (i) extremely high GSC interdependencies between US and Chinese companies in IT industry; (ii) large number of intermediate inputs with a high input supplier concentration (low degree of diversification) due to highly specialised inputs and/or cost advantages; (iii) vulnerability to systemic exposure of IT sector firms to common GSC shocks due to a large fraction of shared input suppliers; and (iv) the total number of actual global input suppliers by accounting for the entire recursive sequence of all the inputs into all the inputs is not exactly known even to these large public companies (known are only the first and second tier input suppliers).

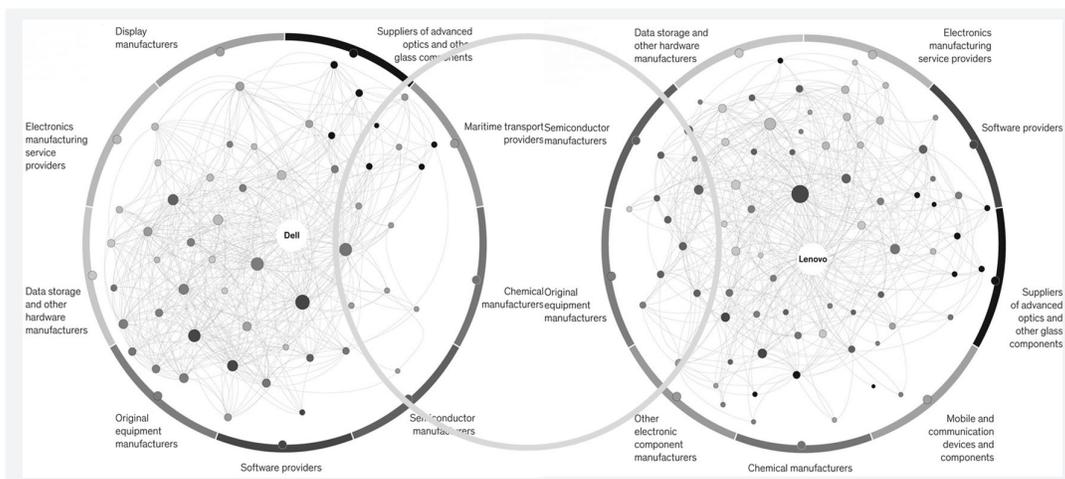

**Figure 1: GSC interdependencies: a US company and a Chinese company in IT industry**
Source: Computed based on www.dell.com/en-us/dt/oem/military.htm; Bloomberg Global Supply Chain Data; and www.dell.com/en-ca/dt/oem/defence.htm data.

Why is the domestic company supply chain foreign exposure important for policy makers? GSCs are characterised by externalities and market failures, which implies that the firm-level equilibrium efficiency-robustness allocation may be inefficient socially (see section 3). Typically, the efficiency-robustness allocation of private sector firms is skewed toward efficiency more than it would be socially optimal, and due to GSC complexity and opaqueness private misjudgements as to how uncertain GSCs actually are may lead to a misperception of the actual vulnerability. An even more important argument for a policy maker attention is given by the increasing deployment of foreign supply dependence as a hybrid threat by adversaries (European Commission 2021). A recent example is Russia's attempt to weaponise energy supplies against Europe.

## 2.2 Macro perspective: Foreign exposure of economies

In the value-added (macro) approach – where the unit of analysis are industries & countries – we look at how production is allocated internationally and how each stage of production contributes to the final product. Combining international trade data with national Input-Output tables yields cross-country or World Input-Output Tables (WIOT). Information contained in these tables allow us to shed light on value-added trade flows across countries and the implied degree to which production processes have become globalised. To measure the international fragmentation of production processes, we rely on insights from the theoretical "macro" literature, which mostly focuses on the development of structural interpretations of the WIOT, with the ultimate goal of constructing reliable tools for counterfactual analysis by acknowledging the relevance of GSCs in the world trade (Antras and Chor 2022).

A variety of metrics has been developed to assess the foreign exposure of a sector or economy as a whole (see e.g. Johnson 2018). For example, the content of value added in final goods, value added in gross exports, positioning in GSCs. Foreign Input Reliance (FIR) measures the sourcing-side exposure of a sector or the entire economy. We use the Inter-Country Input-Output (ICIO) data from the OECD to compute FIR for G7 economies and China in 2019 (the most recent available data). The computed bilateral FIR corresponds to the share of foreign sources used as intermediate inputs into domestic production. Table 1, panel (a) reports row nations' reliance on inputs from column nation for manufacturing production. Cell shades are indexed to share sizes; darker shades indicate higher FIR (more import-dependent).[4] For example, 11.8 in the row for Canada (CAN) and the column for China (CHN) indicates that 11.8% of Canadian manufacturing production was made using inputs sourced directly and indirectly from China. The global dominance of China in intermediate input trade can be seen by the fact that CHN column is shaded primarily in dark. The fact that the CHN column is relatively dark indicates that China is an important supplier of inputs to manufacturing industries of all analysed G7 economies. It is also worth noting the asymmetry between the USA manufacturing production's reliance on Chinese inputs, 9.9%, and China's manufacturing production's reliance on US inputs, 3.7%.

---

[4] The matrix diagonal elements are suppressed, as we are interested in foreign inputs and foreign exposure. The diagonal elements would show a nation's input reliance on itself - both in terms of direct domestic sourcing and indirect sourcing through the re-import of previously exported inputs.

| (a) | USA | CAN | GER | GBR | FRA | ITA | JPN | CHN | ROW |
|---|---|---|---|---|---|---|---|---|---|
| USA |  | 5.4 | 1.8 | 1.0 | 0.7 | 0.8 | 2.1 | 9.9 | 13.0 |
| CAN | 32.5 |  | 2.1 | 1.5 | 0.9 | 0.9 | 2.0 | 11.8 | 21.1 |
| GER | 4.6 | 0.5 |  | 3.2 | 4.7 | 3.8 | 1.6 | 6.9 | 42.0 |
| GBR | 6.2 | 1.4 | 6.9 |  | 4.1 | 2.6 | 1.3 | 7.7 | 29.5 |
| FRA | 5.6 | 0.7 | 10.1 | 3.8 |  | 4.7 | 1.2 | 6.4 | 35.3 |
| ITA | 3.5 | 0.5 | 8.9 | 2.6 | 5.8 |  | 0.9 | 7.6 | 39.6 |
| JPN | 4.1 | 0.7 | 1.3 | 0.7 | 0.6 | 0.4 |  | 10.7 | 26.0 |
| CHN | 3.7 | 0.8 | 1.7 | 0.6 | 0.7 | 0.6 | 3.2 |  | 24.6 |

| (c) | USA | CAN | GER | GBR | FRA | ITA | JPN | CHN | ROW |
|---|---|---|---|---|---|---|---|---|---|
| USA |  | 3.2 | 1.0 | 0.8 | 0.7 | 0.4 | 1.3 | 5.6 | 9.6 |
| CAN | 31.9 |  | 0.8 | 1.3 | 0.6 | 0.3 | 1.7 | 10.8 | 9.3 |
| GER | 7.1 | 0.8 |  | 3.8 | 5.1 | 4.2 | 1.6 | 10.0 | 41.0 |
| GBR | 7.0 | 0.9 | 4.7 |  | 2.9 | 2.1 | 1.2 | 5.5 | 25.8 |
| FRA | 5.2 | 0.7 | 8.4 | 3.9 |  | 5.1 | 1.4 | 8.0 | 33.1 |
| ITA | 5.9 | 0.7 | 6.7 | 2.6 | 4.6 |  | 1.3 | 5.4 | 31.9 |
| JPN | 5.7 | 0.6 | 1.1 | 0.6 | 0.5 | 0.4 |  | 14.4 | 16.8 |
| CHN | 8.0 | 0.8 | 1.3 | 0.9 | 0.7 | 0.7 | 2.8 |  | 15.7 |

| (b) | USA | CAN | GER | GBR | FRA | ITA | JPN | CHN | ROW |
|---|---|---|---|---|---|---|---|---|---|
| USA |  | -1.4 | -0.6 | -0.5 | -0.2 | -0.2 | -1.8 | 6.0 | -3.9 |
| CAN | -1.1 |  | -0.2 | -0.8 | -0.2 | -0.1 | -1.6 | 6.1 | 2.0 |
| GER | 0.7 | -0.2 |  | -0.5 | -0.3 | -0.1 | -0.3 | 4.9 | 5.0 |
| GBR | 1.0 | 0.1 | 0.4 |  | -0.7 | -0.4 | -0.7 | 4.9 | 0.1 |
| FRA | 1.4 | 0.1 | 1.1 | -0.2 |  | -1.1 | -0.5 | 4.0 | 0.2 |
| ITA | 0.4 | -0.1 | 0.1 | -0.4 | -0.8 |  | -0.3 | 5.0 | 3.9 |
| JPN | 0.4 | 0.1 | 0.1 | 0.1 | 0.1 | 0.1 |  | 5.6 | 5.7 |
| CHN | -1.3 | -0.1 | -0.7 | -0.2 | -0.4 | -0.4 | -6.1 |  | -8.2 |

| (d) | USA | CAN | GER | GBR | FRA | ITA | JPN | CHN | ROW |
|---|---|---|---|---|---|---|---|---|---|
| USA |  | -0.5 | 0.1 | 0.1 | -0.1 | -0.3 | 3.8 | 1.4 |
| CAN | -17.6 |  | -0.1 | 0.1 | -0.1 | -0.2 | -1.1 | 8.3 | -0.2 |
| GER | -1.5 | 0.1 |  | 0.3 | 0.1 | -0.8 | 0.1 | 6.5 | 6.0 |
| GBR | -0.8 | 0.1 | 0.1 |  | -0.4 | -0.8 | -0.1 | 3.9 | 2.1 |
| FRA | -0.1 | 0.1 | 1.2 | -0.2 |  | -0.9 | -0.1 | 5.7 | 4.0 |
| ITA | 0.3 | 0.1 | 1.8 | -0.1 | -0.1 |  | 0.1 | 3.9 | 6.6 |
| JPN | -2.3 | -0.2 | 0.1 | -0.1 | -0.1 | -0.1 |  | 7.5 | 3.3 |
| CHN | -3.9 | -0.8 | -0.4 | -0.5 | -0.5 | -0.6 | -3.2 |  | -0.8 |

**Table 1: Panel (a): Foreign Input Reliance in 2019 (FIR, %); Panel (b): Change in Foreign Input Reliance between 2000 and 2019 (ppt); Panel (c): Foreign Market Reliance in 2019 (FMR, %); Panel (d): Change in Foreign Market Reliance between 2000 and 2019 (ppt).**
**Source: Authors' computations based on Inter-Country Input-Output (ICIO) Tables http://oe.cd/icio. Notes: ROW denotes the Rest of the World.**

Next, we investigate how Foreign Input Reliance has changed during the last two decades by comparing FIR in 2019 with FIR in 2000. Table 1, panel (b) reports change in row nations' reliance on inputs from column nation for manufacturing production between 2000 and 2019. Darker-shaded cells indicate larger changes in FIR. For most countries the bilateral FIR matrix with respect to China was considerably larger in 2019 than it was in 2000. In panel (b), the figures in the China column are all positive and all significantly different from zero, indicating that the G7 industries' input dependence on China has increased. In contrast, the figures in the USA column are small, mostly, under 1ppt, and some figures are even negative (e.g. China). Most of the panel (b) entries for other countries are negative. Overall, the reliance of G7 economies on Chinese inputs has increased substantially between 2000 and 2019, whereas the opposite is observed for Chinas reliance on inputs from G7 (last row in Table 1, panel (b)).

Industries and countries participating in GSCs are exposed also to sales-side shocks. Therefore, it is likewise important to understand domestic industries' foreign dependence on the output side. Conceptually similar to the FIR index - which measures countries' total reliance on foreign production on the sourcing side - the Foreign Market Reliance (FMR) index measures countries' reliance on foreign markets on the sales side. Table 1, panel (c) reports row nations' total input sales to column nations' manufacturing industries for G7 economies and China in 2019, again based on the Inter-Country Input-Output (ICIO) data from the OECD. As before, cell shades are indexed to share sizes; darker shades indicate higher bilateral FMR (more foreign market-dependent). Overall, the G7 economies' foreign market exposure with respect to China is high (higher than the bilateral foreign exposure between most G7 country pairs). Second, the global importance of the USA and China stand out from the rest, as the respective columns are primarily shaded dark. However, the bilateral US-China asymmetry is less marked and reversed since China's sales-side reliance on the US is 8.0% while that of the US on China's market is only 5.6%.

Finally, as for the input sourcing side, we also compute the change in Foreign Market Reliance between 2000 and 2019. Table 1, panel (d) reports change in row nations' total input sales to column nations' manufacturing industries, 2019 vs. 2000. Dark-shaded cells indicate large FMR decreases or increases. Overall, panel (c) suggests that the G7 economies' FMR has been further increasing with respect to China during the last two decades. These findings apply both to the input sourcing side as well to the sales side. Given that the foreign exposure is an inverse measure of the domestic industries' resilience and robustness (see trade-off in Figure 1) with respect to GSC shocks, our results imply that the increasing dependence on intermediate inputs from China and market sales in China may contribute negatively to the G7 economies' vulnerability.

## 2.3 Increasing Global Supply Chain vulnerabilities

Because of a widespread production outsourcing, off-shoring and often insufficient investment in resilience in absence of robustness-promoting policies, many global production networks have become excessively complex and fragile (Baldwin and Freeman 2022). The GSCs of 2020s are efficient but brittle – vulnerable to breaking down in the face of a pandemic, a war or a natural disaster. These developments are important to understand, as the increasing fragility of GSCs may have implications for the vulnerability of critical sectors and essential services as well as implications for the entire Alliance's security and defence.

In the absence of systemic shocks to GSCs, the foreign input and output dependence may not be critical. In reality, however, all production structures entail uncertainty, whereby sourcing inputs from abroad exposes domestic activity additionally to foreign shocks, making globally fragmented production structures more vulnerable than locally organised production processes. There are at least three propagation channels exposing the domestic activity to GSC shocks: the costs and effects of delinking; the propagation of micro shocks into macro shocks; and GSCs amplify the trade impact of macro shocks (Antras and Chor 2022).

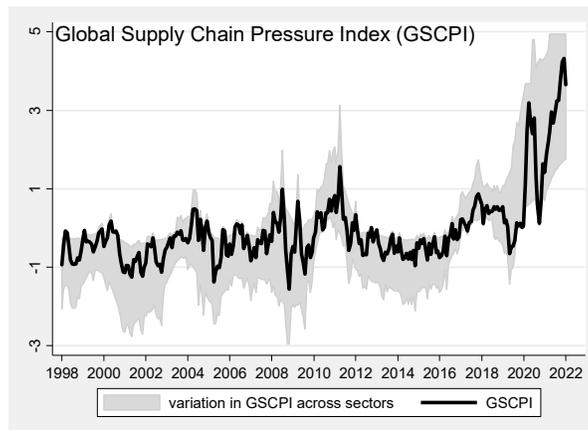

**Figure 2: Global Supply Chain Pressure Index (GSCPI) 1998-2022**
**Source: Computed based on www.newyorkfed.org/research/policy/gscpi**

Different metrics and indices have been developed to monitor and track the state of GSCs. The Global Supply Chain Pressure Index (GSCPI) is one of the most robust indices; it is being deployed by the Federal Reserve Bank of New York. GSCPI measures a common factor of several cross-country and global indicators of supply chain pressures (e.g., delays in shipments and delivery times and shipping costs after purging these from demand measured by new orders). As illustrated in Figure 2, the GSC pressure is at historically high levels since 2020, which is signalling an escalating probability of GSC ruptures.

The GSC-disruption-caused losses are escalating and the frequency and intensity of hybrid threats is increasing, particularly during the recent years (European Commission 2021). Also the World Economic Forum (2021) is explicit about the increasing vulnerability of GSCs to shocks: "The increasing frequency of supply-driven disruptions – ranging from global pandemics and the climate crisis to cyber threats and geopolitical tensions – combined with an ever intensifying set of demand-driven disruptions – including the rise of new consumer channels, pent-up demand and a fragmented reopening of the global economy – will continue to destabilise global value chains." GSC ruptures create uncertainties, some of which can be attributed to GSC risks whereas others to ambiguity; a distinction is important, as they may require differentiated policy responses.

## 2.4   Accelerating the evidence-base

Both the firm-level and aggregate stock-taking exercises of where allies are in terms of the GSC-foreign reliance provide valuable insights to decision makers about the true supply chain pressures and vulnerabilities via foreign input and output exposure. However, the collection of these data takes time, which typically lasts several years and hence implies delays in the evidence base to respond to shocks. In the case of firm-level evidence, the necessary data are collected via numerous sequential company surveys following the entire recursive upstream and downstream sequence of all the input providers of all input inputs. In the case of macro-level aggregate evidence, the time lags are even larger, as national account data, international trade statistics and input-output tables are usually published with a backlog of around three to four years. As a result, the retrospective evidence does not allow to react pro-actively to dynamically changing external threats.

New technologies and advances in big data offer new opportunities to monitor and trace GSCs and locate arising bottlenecks in real time. For example, blockchains allow to record and access real-time data, avoiding unnecessary delays in decision maker response to dynamically changing and intensifying hybrid threats, such as the recent attempts by Russia to weaponise energy supplies against Europe. The content stored on the blocks – and the management of the stored data performed by the various participants – can be securely controlled depending on how the blockchain is configured (Ciaian et al. 2021). There are several types of distributed ledgers that can provide safe real-time solutions for recording, tracking and securing of sensitive data. For example, public blockchains allow anyone to access them; private blockchains are only open to selected users; whereas permissioned blockchains are a hybrid of public and private blockchains where anyone can access them as long as they have permission from the administrators to do so. In the context of GSCs in critical sectors, blockchains can be designed with limited access to designated actors along the Alliance's defence chain.

A private blockchain is the most restrictive distributed ledger that operates as a closed database secured with cryptographic concepts and the organisation's needs. Only those with permission can run a full node, make transactions, or validate/authenticate the blockchain changes. For example, in a private blockchain of a critical defence sector the participation at the network would be only through an invitation where their identity or other required information is authentic and verified. The validation would be done by the network operator(s) or by a clearly defined set protocol implemented by the network through smart contracts or other automated approval methods. Hyperledger is one such private blockchain frameworks – already used by many enterprises and medium-sized companies within the Alliance – that could be deployed within a relatively short period of implementation for the management and secure sharing of sensitive GSC data in critical sectors (Ravi et al. 2022). It was initiated as an open-source distributed ledger by the Linux Foundation in 2016. The current Hyperledger release offers a modular, scalable and secure framework for real-time transactions, compatible with smart contract technology and secure sharing of sensitive data. The key advantage of deploying a private blockchain for the management and security of sensitive GSC data in critical sectors would be that by reducing the focus on protecting user identities and promoting security of data, efficiency and immutability (the state of not being able to be changed by adversaries) is prioritised (Ravi et al. 2022). A second important advantage would be a continuous access to real-time data, allowing to avoid unnecessary delays in decision maker response to GSC shocks and potential Alliance's vulnerabilities.

## 3.0   SIMULATION-BASED DECISION SUPPORT: A VIEW FORWARD

In the previous section, we have taken a retrospective snapshot of foreign input reliance. In this section, we look forward – what can decision makers undertake proactively to enhance the robustness/resilience of supply chains? We aim to understand the potential consequences of extreme events – like natural disasters, pandemics, hybrid and/or military aggression – on GSCs and the participating upstream producer resilience and robustness under alternative organisational structures. Model-based simulations will help us to answer this question. We undertake GSC stress tests by simulating counterfactual shock scenarios under alternative uncertainty setups. According to our simulation results, the optimal input sourcing diversification depends on the GSC integration and the nature of shock uncertainty.

## 3.1 Modelling framework

Inspired by the theoretical GSC literature (for an overview, see Antras and Chor 2022), which is largely concerned with developing tools to solve the complex problems that firms face when designing their optimal global production decisions – forward GSC participation, backward GSC participation, centralised versus lead-firm approaches – the underlying conceptual framework (see Annex A.1) is a parsimonious supply chain model based on Antràs and de Gortari (2020) and Jiang et al. (2022). For the sake of brevity, we abstract from many important GSC-related decision of firms highlighted in recent literature, such as staggered GSC participation, buyer-supplier matching, or relational nature within GSCs.[5] In our stylised supply chain model, we focus on key forward GSC participation and backward GSC participation decisions of firms to study aggregate shocks, and investigate how the GSC diversification and robustness changes under risk and ambiguity.

The model GSC consists of two production stages that need to be performed sequentially by two types of firms: small intermediate good upstream suppliers and one globally sourcing downstream firm producing a final demand good. Production in the initial stage, $n = 1$, only uses labour, while the second stage of production, $n = 2$, combines labour with the intermediate input good produced in the first stage. There are two countries, $J = 2$: the East and the South, which differ in the probability that an aggregate shock hits. Generally, the countries may differ in terms of unit production costs, $c$, productivity, $\omega$, trade costs, $\tau$, prices, $p$, and face other asymmetries. In counterfactual simulations, however, to introduce the modelling framework and gain the intuition of the main mechanics, we employ a version with symmetric prices and costs across the two locations. Time $t$ occurs in discrete steps; the discrete time version of the model is convenient to characterise and compare the equilibrium solutions.

Presume there is a sufficiently large number of intermediate input firms at the start of period $t$. Each upstream firm produces a single and unique intermediate good that is different from other intermediate input producers. At the beginning of every period, each intermediate input firm faces a location decision to choose one of the two locations where it sets up production, whereby locations may be subject to aggregate shocks. The location decision minimises cost, $c$, by choosing the least-cost path of production. Given that the intermediate input firms do not care about their survival, this amounts to choosing locations $l_j^1$ and $l_j^2$ to minimise cost $c(l_j^1, l_j^2)$. The production of the intermediate good is uncertain. In each period $t$, one of the two locations may suffer an aggregate shock with arrival probability $\varepsilon$; hence, there is no aggregate shock with probability $1 - \varepsilon$. Conditional on such a shock occurrence, and before the production occurs, all intermediate good firms in the East or South perish with probability $\zeta$ or $1 - \zeta$, respectively. In counterfactual simulations, we presume that $\zeta \gg 1/2$, implying that the East is riskier than the South. In the following period, $t + 1$, the intermediate input production takes place by the surviving firms. Between periods, the number of intermediate input producers can grow with a growth rate A, which depends on the total number of surviving input producers in the world. The new entering intermediate input producers are distributed across locations according to the existing shares of surviving input firms in each location. The evolution of intermediate input firms depends on the aggregate shock realisations.

In designing a cost-optimal organisational structure, the downstream firm chooses the ownership structure, i.e., vertical integration (centralised) versus outsourcing (decentralised). It aggregates intermediate goods from all the input suppliers and manufactures a final demand good. The sophistication of the final demand good depends on the number of intermediate inputs used in its production. The final demand good is more desirable, the more inputs are used in its production – the quality increases with the number of inputs. The revenue from selling the final demand good is linear in the number of intermediate inputs it contains; hence, the globally sourcing downstream producer wants to use as many parts as possible to produce and sell the final demand good internationally. Hence, the downstream firm cares about intermediate input supplier survival. The long-term benefit of the downstream firm is the continuation value, $v$. The expected value of continuation is the differences between the revenue – which depends on the number of surviving firms – minus the cost. At least one intermediate input is needed to produce the final demand good. All the revenue from the downstream firm is transferred to the intermediate input firms, which is equally shared among all input producers (surviving firms and new entrants) to pay the cost before starting production in the next period (the downstream firm has zero profits).

---

[5] For theoretical foundations of these channels of GSC adjustment, we refer to Antras and Chor (2022).

In counterfactual simulations, we assume that the price per produced intermediate input, $p$, is constant and independent of production and the state of the world; the small intermediate input producers are price takers. Given that prices are fixed, they cannot adjust after a shock and there is no way of compensating the intermediate input producers that locate themselves in a region. Note that the prices are fixed is as long as the investment horizon, which is one period. At the end of every period, the upstream firms can relocate between regions without cost (there are no adjustment/switching costs in the model).

The model incorporates the well-known trade-off between efficiency and robustness already mentioned in section 1 (Lettau and Ludvigson 2003). In the pursuit of efficiency, a supply chain could become vulnerable to aggregate shocks. On the other hand, a supply chain with a greater robustness to shocks has to sacrifice efficiency during normal times (in absence of shocks). In the model, there is both a marginal benefit and cost of diversification. Profit maximising firms (integrating in GSCs or not) aim to ensure a certain resilience/robustness of the production process while keeping the cost optimisation in mind.[6] In the context of the downstream firm's input sourcing decisions, a key trade-off in both resilience and robustness decisions at the firm level involves diversification of uncertainty versus lower cost and higher quality inputs. The trade-off between the uncertainty (in form of risk and ambiguity) that comes with GSCs (vertical axis) and the rewards (horizontal axis) is illustrated in Figure 3. The solid line represents the uncertainty-reward frontier; everywhere on this line the firms' willingness to substitute one unit of uncertainty for reward is constant. Moving on this line starting from the origin we can think as sourcing from fewer locations and more specialised upstream suppliers. Uncertainty is assumed to increase as the downstream firms concentrate production of inputs in the single cheapest location. Diversification of input sources reduces risk and ambiguity but at a diminishing rate. Solving the underlying mathematical model, the equilibrium solution is found at the tangency of the indifference curves and the uncertainty-reward frontier, represented by point $P$ in Figure 3.

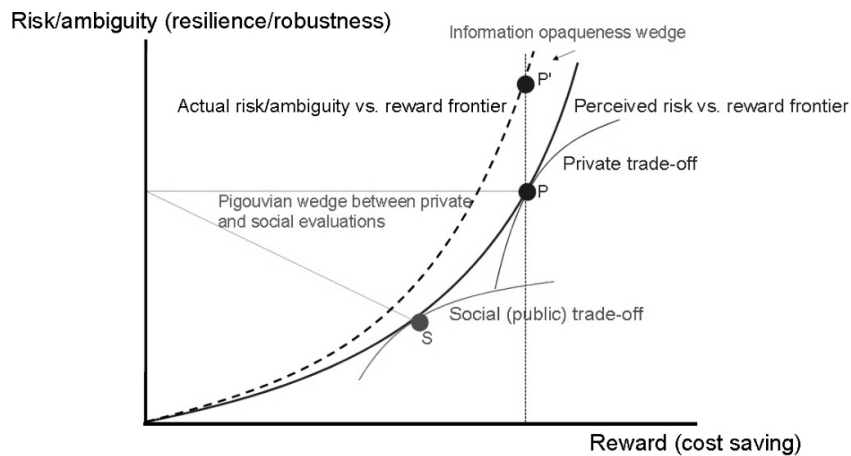

**Figure 3: Firms' efficiency-robustness trade-off, externalities and market failures**
**Source: Based on Lettau and Ludvigson (2003) and Baldwin and Freeman (2022)**

Although optimal from the perspective of a single firm, the equilibrium efficiency-robustness outcomes may be inefficient socially in the presence of externalities and market failures.[7] First, social evaluation of the uncertainty-reward trade-off may put a greater stress on uncertainty than private evaluation. Private companies may accept more risk/ambiguity (resilience/robustness) for any given level of reward compared to the society, which usually cares relatively more about risk/ambiguity (resilience/robustness). The indifference curve shapes (and position in Figure 3) reflect that private sector firms would agree with more risk/ambiguity for any given level of reward (curve 'Private (decentralised) trade-off'), but the public cares relatively more about risk/ambiguity (curve 'Social (centralised) trade-off'). In equilibrium, the society is desiring a lower level of uncertainty, point $S$, than the private sector, point $P$ in Figure 3. This wedge between the centralised and decentralised evaluation for risk ('Pigouvian wedge') is an externality that is not internalised by economic actors in their optimisation decisions leading to a market failure. As has become evident during the recent peaks in GSC ruptures (see Figure 2), markets for medical supplies share features of the public-private wedge, as do other 'strategic' inputs such as semiconductors.

---
[6] Lettau and Ludvigson (2003) refer to the risk-reward trade-off as a decision process of firms, which typically care about both uncertainty (i.e. they value resilience/robustness), as well as the reward from cost savings.

[7] See, e.g. Turvey (1963) for the underlying concept.

The equilibrium efficiency-robustness outcome is likely to be socially suboptimal also in markets where a collective action problem creates information asymmetries that force companies to act without a full information (Baldwin and Freeman 2022). In 2020s, GSCs are characterised by complexity and non-transparency. As shown in section 2.1, even large, sophisticated companies do not know all their suppliers and the suppliers of their suppliers, and even seemingly 'purely domestic companies' might not appreciate being part of a global network. The general lack of firms' understanding of where their input-inputs originate from – supply chain opaqueness – implies that companies may be sub-optimally making decision with respect to the risk-reward trade-off and misaligning their input sourcing and output supplies. Russia's war on Ukraine and the implications on global food and energy supplies visibly demonstrate how this lack of information about where domestic company inputs and input-inputs are sourced from can result in private misjudgements about the actual vulnerability of GSCs. In Figure 3, the supply chain opaqueness-caused market imperfections are shown on curve 'Actual risk vs. reward frontier', which is above the 'Perceived risk vs. reward frontier' curve. We refer to the gap between the two curves as the 'Information opaqueness wedge' in Figure 3.[8] Since GSCs are highly interwoven and generally not fully contained within the boundaries of a single firm (Figure 1), information about them resembles a public good features. This information is costly to collect, cheap to share, and provides value to many.

### 3.2 Simulation scenarios of GSC robustness

In order to study how a GSC's decision of input sourcing diversification versus efficiency changes in the presence of uncertainty and what are the implications for supply chain robustness, we set up a number of counterfactual scenarios, and simulate alternative shock realisation possibilities, setups of shock expectations and arrangements of the GSC organisational structure. The key modelling assumptions in the counterfactual scenario construction are summarised in Table 2.

First, we study various aggregate shock scenarios with respect to the conditional probability that the extreme event occurs. (i) In the risk-free scenario, the value of the conditional probability that the aggregate shock occurs is known. In line with Knightian uncertainty, there are two categories of imperfectly predictable events, which involve different choices: risky events and ambiguous events. (ii) In the GSC risk scenario, the distribution of the conditional probability that the aggregate shock affects the location is known (the shock is distributed according to some a priori known distribution; in our simulations we assume that the distribution is uniform). (iii) In the GSC ambiguity scenario, the shock exhibits a bounded uncertainty and the distribution of the conditional probability that the aggregate shock affects the location is unknown; all firms know that a shock can occur in one of the regions but they do not know the underlying distribution. These differences in the conditional probability that the shock hits the particular location are summarised in rows of Table 2.

Second, we investigate two different organisational structures of the GSC: a decentralised GSC with upstream outsourcing versus a vertically integrated GSC.[9] In the upstream outsourcing GSC organisational structure, the inputs-producing upstream firms individually decide their location and the downstream firm purchases the surviving suppliers' output. The intermediate input producers are small; they do not take into account the impact their decision has on the location decision of others. Due to the externality explained in Figure 3, decentralised supplier's location decisions may not be socially optimal (when the probability of global survival is taken into account). This wedge between private and social allocations comes from the pricing system's inability to compensate input suppliers properly for moving into the South. Given that there is no cost of switching between locations, the intermediate input firms are solving a static problem – the continuation value is the same for all firms. Input producing firms are maximising the expected value of East versus South. In the integrated GSC, the globally sourcing downstream firm is vertically integrated with intermediate input suppliers and hence can determine their location. The input suppliers being its subsidiaries, the downstream firm can centralised choose all its suppliers' locations, thereby internalising the location decision. In the input supplier allocation decision, the downstream firm takes into account the overall distribution of intermediate input firms, and the expected value of continuation in all states of the world. By internalising the externality, the globally sourcing downstream firm aims to ensure all its input suppliers survive. These differences in the integration arrangements of the GSC are visible when comparing rows of 2-4 with rows 5-7.

---

[8] The incomplete information regarding upstream suppliers supports the Fréchet distribution of productivity at the chain level that we specify in the model (see Annex A.1).

[9] In Antras and de Gortari (2020) the vertically integrated scenario is referred to as a lead-firm problem.

|  | No aggregate shock | Aggregate shock in East | Aggregate shock in South |
|---|---|---|---|
| Risk-free, outsourcing | Externality, known probability, no shock realised | Externality, known probability, shock realised in East | Externality, known probability, shock in South |
| GSC risk, outsourcing | Externality, known distribution no shock realised | Externality, known distribution shock realised in East | Externality, known distribution shock in South |
| Ambiguity, outsourcing | Externality, unknown distribution, no shock realised | Externality, unknown distribution, shock realised in East | Externality, unknown distribution, shock in South |
| Risk-free, vertical integration | Internalised externality, known probability, no shock realised | Internalised externality, known probability, shock realised in East | Internalised externality, known probability, shock in South |
| GSC risk, vertical integration | Internalised externality, known distribution no shock realised | Internalised externality, known distribution shock realised in East | Internalised externality, known distribution shock in South |
| Ambiguity, vertical integration | Internalised externality, unknown distribution, no shock realised | Internalised externality, unknown distribution, shock realised in East | Internalised externality, unknown distribution, shock in South |

**Table 2: Simulation scenarios, key assumptions about shocks and uncertainty**

Finally, the location dimension allows to investigate how the evolution of intermediate input firm population and hence the supply chain's input reliance depends on the aggregate shock occurrence: (i) there is no aggregate shock (with probability $1 - \varepsilon$); (ii) the aggregate shock hits the East in period 10 and all firms in the East perish (with probability $\varepsilon * \zeta$); (iii) the aggregate shock hits the South in period 10 and all intermediate input producers in the South perish (with probability $\varepsilon * (1 - \zeta)$). These three shock possibilities are summarised in the columns of Table 2.

### 3.3  Simulation results: ex-ante robustness

Our parsimonious model allows, among others, to analyse the survival probability of a GSC, and complex linkages between vertical integration and uncertainty expectations under different organisational structures. Simulation results are summarised in Figures 4-6, which show how the input sourcing diversification of a GSC changes in the presence of uncertainty. The black dotted line describes the intermediate input firm location decisions and their survival in the South, the grey dashed line in the East, whereas the solid red line reports the total number of operating intermediate input firms in each specific GSC setup.

In the decentralised upstream outsourcing setup (left panels in Figure 4), given our assumption of lower shock probability in the South, it is the preferred location for all intermediate input firms. The value of locating in the South is always larger than the value of locating in the East; the continuation value for each input producer, conditional on having survived the aggregate shock, is independent of location. Simulation results presented in the left panels in Figure 4 suggest that the fragmented GSC results in a corner solution (all firms locate themselves in the South). While the individual intermediate input producers maximise their own efficiency (productivity), indirectly they expose the downstream firm to an aggregate shock to the South. In the vertically integrated GSC setup (right panels in Figure 4), the downstream firm maximises survival, hence, it internalises the survival externality. It is willing to allocate intermediate input production facilities also in the East to insure production against an aggregate shock in the South - it internalises the survival probability and diversifies the input sourcing. In the right panels of Figure 4, the dashed grey lines (number of firms in the East) are above zero in all scenarios, implying that the integrated GSC results in an internal solution - a share of intermediate input firms larger than zero in the East.

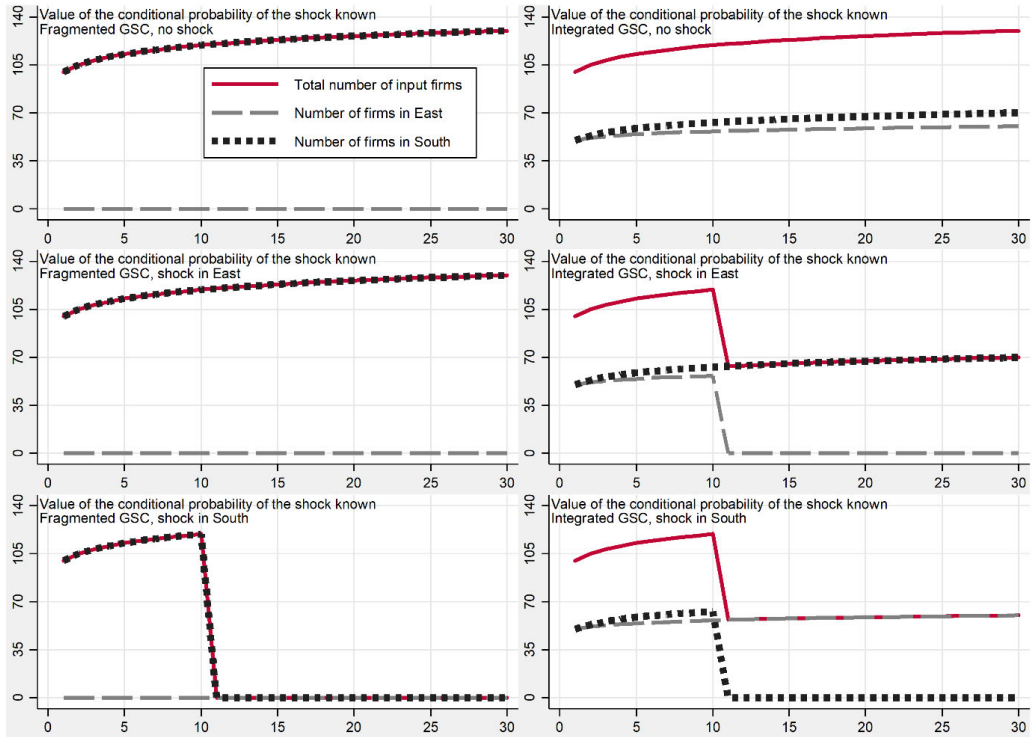

**Figure 4: Simulation results with known value of the conditional shock probability.
Notes: Y-axis measures the number of surviving intermediate input producers; X-axis refers to time periods. Upstream outsourcing is in left; centralised GSC in right panels**

Next, consider the GSC risk scenario in Figure 5, where the distribution of the conditional probability that the aggregate shock affects the location is known. Without loss of generality, in our simulations we assume that the distribution is uniform. In the upstream outsourcing setup (left panels in Figure 5), the risk-neutral intermediate input firms are maximising the expected value of East versus South and given our assumptions South dominates for all intermediate input producers. Given that upstream firms are risk-neutral and the East is assumed riskier, the worst case for each individual input-producer still implies that in terms of the continuation value the worst case in the South is better than the worst case in the East. As in the risk-free scenario, also in the GSC risk scenario the intermediate input firms do not take into account the survival probability of the downstream firm. In contrast, the risk-averse downstream firm exhibits an allocation behaviour that takes into account all the sources of risk - a desire for diversification. The diversification is obtained as in a traditional expected utility maximisation problem by either increasing the variance or the risk aversion to infinity. The downstream firm faces a trade-off between instantaneous profits (what the individual intermediate good producers maximise) and the probability of survival. The left panels in Figure 5 also suggest that the final good producer's optimal allocation of intermediate input activity to a particular location depends on the number of active (survived) suppliers.

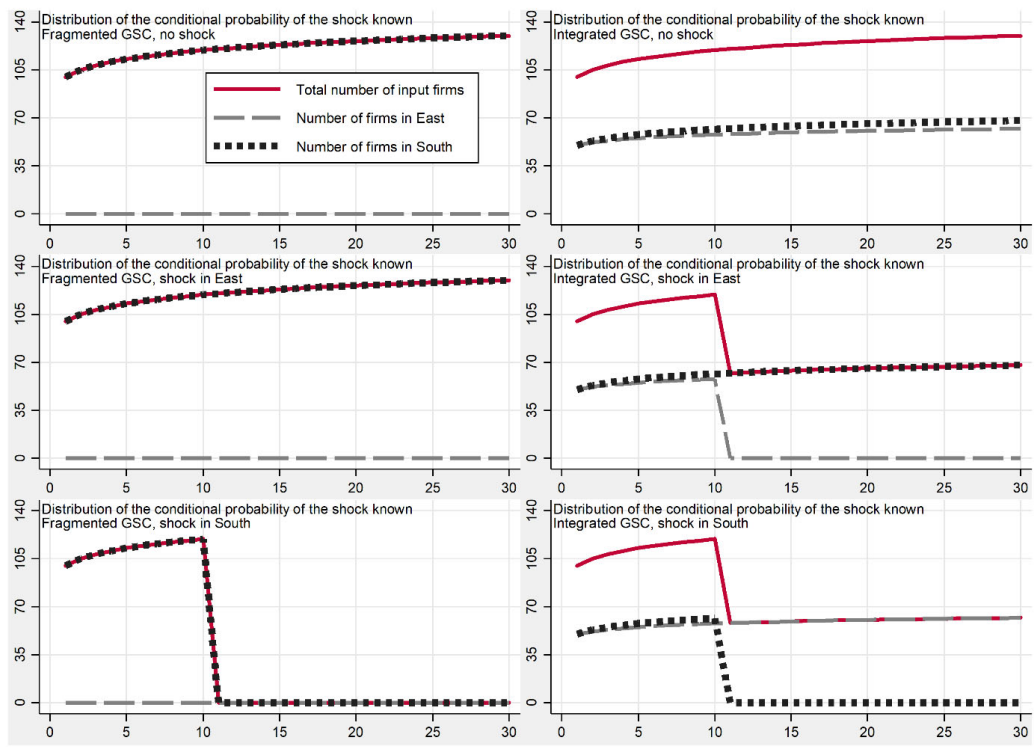

**Figure 5: Simulation results with known distribution of the conditional shock probability. Notes: Y-axis measures the number of surviving intermediate input firms; X-axis time periods. Upstream outsourcing is in left; centralised GSC in right panels**

In the GSC ambiguity scenario, the intermediate input producers and the downstream firm are averse to ambiguity and hence maximise the expected profit assuming the worst-case value of the conditional probability that the aggregate shock affects the region of firm's location. Ambiguity-averse downstream firm's optimal allocation of intermediate input suppliers to the South is independent of the number of surviving intermediate input producers. It does not matter how much uncertainty faces the East relative to the South, there is a level of ambiguity for which the downstream firm allocates half the intermediate input activity in the East. According to the simulation results reported in the right panels of Figure 6, a robust supply chain is one in which the survival probability is maximised, and where production will continue even under the most demanding circumstances.

A robustly optimal allocation of intermediate input producers is uniform across locations; in our example, half of the intermediate input producers locating in the East and half in the South. This implies that in the presence of an aggregate shock – independently where it occurs – half of the input producers would perish, half would survive. The flows and costs are identical, therefore, the expected continuation value is independent of the true shock realisation. A robustly optimal allocation is the optimal strategy, when the GSC survival is important, and the downstream firms do not know the distribution of the shock they are facing and therefore need to be prepared for the worst. The simulation results presented in the right panels of Figure 6 summarise strategies that reduce the differences over all possible states of nature. The downstream firm's robustly optimal strategy guaranteeing the survival of the supply chain is comparable to insights from the behavioural finance literature, where in many cases agents tend to choose the robust survival maximising strategy (Lettau and Ludvigson 2003).

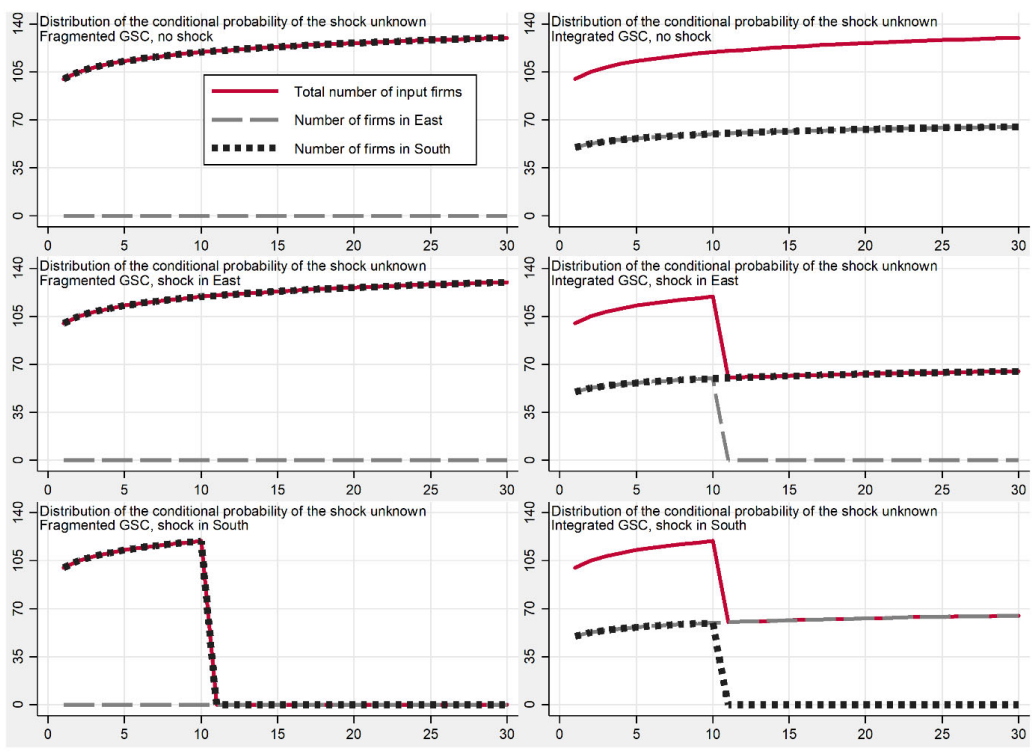

**Figure 6: Simulation results with unknown distribution of the conditional shock probability. Notes: Y-axis measures the number of surviving intermediate input firms; X-axis time periods. Upstream outsourcing is in left; centralised GSC in right panels**

Comparing the left panels in Figures 4-6, we may conclude that under upstream outsourcing, the optimal firm choices are independent of the nature of the shock. In a decentralised GSC, individual input producers will choose a corner solution (e.g., locating exclusively in the South) irrespectively of whether they are facing risk or ambiguity – exposing the downstream firm to an aggregate shock to the South. Whereas individual intermediate input firms maximise efficiency (productivity), the downstream firm maximises survival. The optimal behaviour of the downstream firm is rather different in a vertically integrated GSC (right panels in Figures 4-6), where the downstream firm's optimisation tends to result in internal solutions – a proportion larger than zero of intermediate input firms in the East). When the downstream firm faces risk, the optimal allocation of the upstream activity is an increasing function of the number of input supplying firms (right panels in Figure 5). When ambiguity is present, the optimal firm allocation solution is an internal and fixed ratio (right panels in Figure 6). Since GSC disruptions in critical sectors may have catastrophic impacts on social welfare, and the probability of such disruptions is not known even approximately, uncertainty and robust decision rules are the appropriate tools for analysis and policy recommendations in sectors such as energy supplies, food and water, and communication.

### 3.4 Pro-resilience policies and policy constraints

Next, we investigate the role of government policies on enhancing the GSC resilience. As discussed in section 2, the sourcing of specialised inputs and cost advantages are among the main drivers of the raise of the international trade in intermediate goods resulting in highly complex globally fragmented production and trade networks. Given substantial intermediate input price differences across world-wide locations (Antras and de Gortari 2020), we simulate a government policy subsidising the diversification of input sourcing and hence enhance resilience in a decentralised global economy with an upstream outsourcing ownership structure and externality.

In order to implement the "not trading long-term security needs for short-term economic interests" trade-off conceptually, we frame it as a constrained optimisation with two binding constraints: a robustness/resilience constraint 'long-term security needs' and a resource mobilisation constraint 'short-term economic interests'. The constrained optimisation targets the baseline resilience while doing as little damage as possible to the society's socio-politico-economic fabric. The robustness constraint ensures that the baseline resilience requirements with respect to foreign input sourcing are fulfilled – the supply chain's capacity is subject to a minimum resilience (floor) constraint. The resource mobilisation constraint implies that governments do not ask the impossible of the domestic economy and society – the supply chain's upstream diversification is subject to maximum efficiency loss (cap) constraint. While private sector firms may be willing to temporarily forgo possible gains or even accept losses, especially when it is in the name of a good cause, profit maximising firms' tolerance of forgoing profits is not infinite and therefore is accounted for in the analysis.

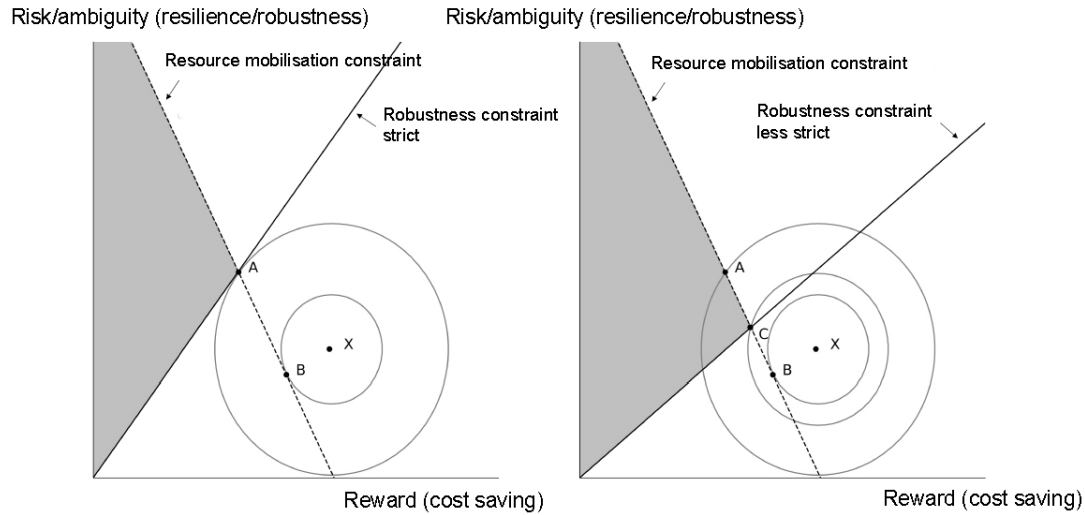

**Figure 7: Efficiency-resilience and constrained optimisation**
**Source: Based on Lettau and Ludvigson (2003) and Baldwin and Freeman (2022)**

Figure 7 shows the key intuition of the constrained policy optimisation problem graphically.[10] The aggregate welfare is represented by indifference curves (circles), with each circle representing a different level of welfare. The maximum welfare in the absence of shocks and policies (and abstracting from other factors such as externalities and market imperfections) is represented by point $X$. The solid line represents the uncertainty-reward frontier (as in Figure 3); everywhere on this frontier the domestic industries' robustness/resilience to GSC shocks is constant. Solving the underlying mathematical model, the equilibrium solution after the implementation of a binding robustness/resilience constraint is found at the tangency of the indifference curves and the uncertainty-reward frontier, represented by point $A$ in the left panel of Figure 7. The other boundary condition is represented by the resource mobilisation constraint - the dashed line in Figure 7. Under these two constraints, the new equilibrium state of the economy would be represented by the solution to the welfare maximisation problem subject to both constraints, which would occur at point $A$ in the left panel. Note that the grey shaded area represents the feasibility region of all possible combinations. Figure 7 also illustrates a scenario with less ambitious resilience goals (right panel). The robustness/resilience constraint is less steep, implying that the new equilibrium is now at point $C$. The level of welfare resulting from these minimum resilience standards (represented by the circle going through point $C$) is not as high as under the optimal resilience strategy (the circle going through point $B$) but it is closer to the optimal than the welfare achieved under the alternative strategy (the circle going through point $A$). The right panel of Figure 7 demonstrates that this strategy represents a more efficient outcome.

---

[10] The solution of the full simulation model is more complex (for details, see Antras and de Gortari 2020 and Jiang et al. 2022).

In line with the Alliance's resilience strategy,[11] we have set up and simulate a baseline scenario and two adverse scenarios. In the risk-free baseline, the value of the conditional probability that the aggregate shock affects the GSC is known. Further, two adverse scenarios with Knightian uncertainties are simulated. In the GSC risk scenario, we simulate 'demanding circumstances' in which the expected GSC shocks have with a priori known distribution. The risk scenario corresponds to the extreme events that are frequent and idiosyncratic (shocks to the firm) studied in the GSC literature (Antras and Chor 2022). In the GSC ambiguity scenario, we study 'most demanding circumstances' in which the expected GSC shocks with an unknown distribution. This scenario is designed to stress-test the GSC in light of the baseline robustness requirements regarding the core functions of continuity of government, essential services to the population and civil support to the military – which must be maintained under the most demanding circumstances. The ambiguity scenario corresponds to the uncertainty analysis with infrequent and aggregate events of Jiang et al. (2022). The simulated GSC shock occurs in 2022, firm responses are plotted until convergence to a steady state (2026). The simulated resilience policy aims to compensate the downstream firms' input sourcing disadvantage from diversifying locations due to global differences in input prices. The policy can be interpreted as an ex-post subsidy to the allocation of an intermediate input production facility in a disadvantaged (from the efficiency point of view) location.

### 3.5    Simulation results: ex-post resilience

The counterfactual results of the three scenarios are summarised in Figure 8. Percentage changes in the number of upstream suppliers are represented by the black solid line. Note that we assume that the initial number of intermediate good suppliers are in equilibrium in the base year (2022) under a decentralised cost-minimising allocation. The two policy constraints – a resilience constraint and a resource mobilisation constraint – are shown by dashed red lines. In counterfactual simulations, the minimum resilience constraint is set at -50% of the initial number of the upstream suppliers; maximum resource mobilisation (efficiency loss) constraint is set at +100% percent, which would imply doubling the number of upstream suppliers. The policy cost – input sourcing diversification subsidy – is captured by red bars. The government diversification subsidy is zero in the risk-free baseline, whereas in the GSC risk and GSC ambiguity scenario the total disbursed amounts are equal. The shaded areas in Figure 8 provide confidence intervals at 90%, 95% and 99% levels.

In the risk-free baseline, the optimal downstream firm strategy implies that the solution to the efficiency-resilience trade-off is highly skewed towards effectiveness. The decentralised optimisation does not internalise the probability of a GSC survival in their optimisation. In the presence of extreme shocks to GSCs, such a downstream firm strategy makes the GSC highly vulnerable and the survival probability is low (top left panel in Figure 8). In the GSC ambiguity scenario, the optimal firm strategy achieves greater resilience through diversification (the steady state in 2026 has around 50% above the initial equilibrium number of intermediate input firms in 2022 in the bottom right panel in Figure 8). The government diversification subsidy makes a substantial impact on firm strategy – by internalising shock ambiguity, ambiguity-averse downstream firms seek to maximise the payoff in the worst-case scenario of a set of potential strategies (bottom right panel). Compared to an efficiency-optimised GSC (zero line on Y axis) in Figure 8, a resilient supply chain with a diversified input sourcing strategy may appear excessively costly in terms of efficiency losses, because the input source diversification and improved resilience are not the most efficient solutions. However, under the 'most demanding circumstances', such a firm strategy makes the supply chain highly resilient and the GSC survival probability is high. In the GSC risk scenario (bottom left panel in Figure 8), the optimal downstream producer strategy reconciles both efficiency and resilience. The intermediate input-sourcing costs are higher than in the risk-free scenario, but lower than in the ambiguity scenario. On the other hand, the increase in the number of upstream suppliers (black solid line) will not be able to achieve the GSC's diversification of the ambiguity scenario though it still will be above the risk-free scenario. The government diversification subsidy contributes substantially to the upstream supplier diversification (the steady state in 2026 is around 40 percent above the initial equilibrium number of intermediate input firms in 2022).

---

[11] www.nato.int/docu/review/articles/2019/02/27/resilience-the-first-line-of-defence/

The optimal downstream firm's strategy under the three simulated scenarios depends on the nature of the expected shocks and the policy framework (subsidies/taxes, maximum resource mobilisation (cap) and minimum resilience (floor) constraints). In the GSC risk scenario, a decentralised strategy implies greater risk aversion, which leads to higher firm incentives for diversification. That is not the case in the GSC ambiguity scenario where resilience cannot be achieved with an infinite risk aversion by firms. According to the conceptual framework and simulation results presented in Figure 8, the diversification implied by centralised decision rules in the ambiguity scenario is both quantitatively and qualitatively different from the diversification that is obtained under the risk-averse firm strategy by either increasing the variance or the risk aversion to infinity (GSC risk scenario). Under the 'most demanding circumstances', the centralised firm strategy implies that resilience can be enhanced with an ambiguity aversion by firms.

The optimal downstream firm's strategy in the GSC risk scenario – illustrated in the bottom left panel in Figure 8 – corresponds to a resilient supply chain, which can optimally deal with risk and recover under demanding circumstances in the medium- to long-run. Policies facilitating reasonable minimum resilience standards to non-critical sectors may ensure both resilience from a security perspective and sustainability from an economic perspective in the medium- to long-run. In the GSC ambiguity scenario – shown in the bottom right panel in Figure 8 – the supply chain can optimally deal with ambiguity, and function under most demanding circumstances. From a policy perspective, implementing such a policy with the highest minimum resilience standards to critical sectors of the economy may be the most GSC-shock- proof and resilient strategy.

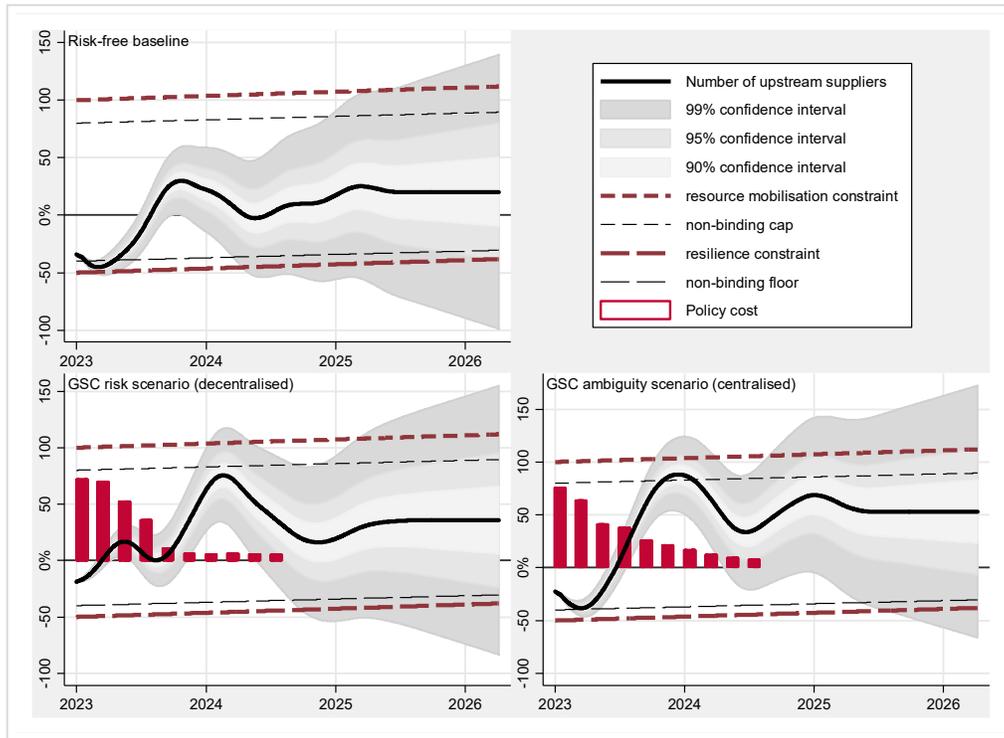

**Figure 8: Simulation results: GSC resilience under alternative shock uncertainties and resilience policies. Notes: Y-axis measures percentage changes in the number of upstream suppliers; X-axis refers to time periods**

Our simulations highlight that the government can effectively align private incentives by imposing minimum resilience standards and by providing a subsidy to the input-sourcing relocation away from riskier locations in the GSCs. Indeed, Japan did so as a response to COVID-19 by setting up a fund to compensate firms that diversify out of China (Jiang et al. 2022). Given differences and the societal sacrifices and resource mobilisation costs across these scenarios, a distinction between 'critical sectors' and 'non-critical sectors' seems paramount to manage the load on domestic producers and possible adverse effects on the tolerability constraint. In the Alliance, the resilience baseline requirements determine the critical sectors and essential services, which must be maintained under the most demanding circumstances.

## 4.0 CONCLUSION

The landscape of hybrid threats is expanding and production processes are increasingly fragmented global across borders. Because of outsourcing, off-shoring and insufficient investment in resilience, many supply chains across the globe have become highly complex and fragile. Different forms of hybrid threats are associated with different types of uncertainties, such as risk and ambiguity. GSC vulnerabilities are important to understand, address and monitor, as the escalating fragility of GSCs may have severe implications for the functioning of critical sectors and essential services, such as energy supplies, food and water, communication networks and transport systems under the most demanding circumstances, as well as implications for the entire Alliance's security and defence. The optimal robustness/resilience strategy will depend on political priorities – not trading long-term security needs for short-term economic – and the political feasibility (robustness/resilience constraint and resource mobilisation constraint). The presented model-based simulations provide an interoperable and directly comparable conceptualisation of positive and normative effects of counterfactual resilience and robustness policy choices under individually optimal (decentralised) and socially optimal (centralised) GSC organisation structures.

A major source of structural problems reducing the resilience and robustness of GSCs is the classical Pigouvian wedge between individual and social evaluations. Given that the society cares relatively more about uncertainty (resilience/robustness) than private sector firms, which prefer more risk for any given level of reward, we demonstrate in simulations that a targeted policy intervention may improve market outcomes in terms of efficiency-robustness. The defence sector is habitually marked by a mismatch of 'supply' of defence services and the 'demand' for defence services. In the defence sector, the Pigouvian wedge between private and social evaluations reducing the resilience and robustness of GSCs is usually addressed via a direct governmental control. A commonly adopted solution in the Alliance is to build in massively redundant 'production' capacity. That is, armies are kept at the ready for years, even decades, without ever being called into combat. Trying to accomplish this resilience via taxes and subsidies, or regulation, may be ineffective and inefficient given the vast cost involved. Instead, a direct public ownership of the military production facilities is the optimal policy choice from an efficiency-robustness perspective.

The current work adds value and contributes along a number of dimensions to the existing modelling and simulation exercises at the Alliance and Member States. First, it integrates several horizontal cross-cutting PMESII elements in one global modelling framework. Uncertainty, externalities and market imperfections – present in many global supply chains – are simulated in counterfactual scenarios under binding socio-politico-economic constraints. Given that GSC disruptions in critical sectors may have catastrophic impacts on the socio-politico-economic fabric, and the probability of such disruptions is unknown, uncertainty and robust decision rules are shown to be the right tools for robustness- and resilience-enhancing policy recommendations. Second, a particular attention is paid to critical sectors, potential vulnerabilities are assessed based on the severity and likelihood of their disruption against a range of stress test scenarios. Data from global Input-Output Tables reveal that in a number of highly specialised industry-country pairs on the sourcing/selling side even a small shock to supply/demand can have major ramifications on the entire socio-politico-economic fabric. To answer the simple question 'where are things made?' comprehensively – as increasingly needed by defence decision makers – one needs to look at foreign input reliance by taking into account the entire recursive sequence of all inputs and all inputs of inputs, not just the first-tier inputs. Third, model-based simulations enable a better understanding of the complexities underlying GSCs and provide a scientific evidence base to a resilience-enhancing decision support. A decentralised GSC and a vertically integrated GSC imply different optimal input sourcing diversification strategies and hence the resilience and robustness.

The results of model-based simulations – as presented in the paper – allow the identification of vulnerability sources and assessment of possible mitigation strategies that could strengthen supply chains in an effective and efficient manner. The decision maker choice of the most suitable strategy in each critical and non-critical sector should depend on the nature of the shocks, source of vulnerability, strategic priorities and resource mobilisation possibilities in the short-, medium- and long-run. Our results have also practical implications and suggestions for decision makers. First, we urge for an Alliance-wide assessment of the key capability areas and economy sub-sectors that the Alliance's security relies upon, including the mapping of critical sectors' vulnerabilities driven by GSCs. The discussed GSC evidence can provide a framework for the necessary evidence base. Second, a stock taking exercise is needed to identify what is directly available within the Alliance (including its strategic partners) to meet the seven baseline resilience requirements under the most demanding circumstances – such as a complete input sourcing cut-off from authoritarian regimes – and what is needed in the short-, medium- and long-run to achieve the baseline resilience requirements. Third, a strategic framework for addressing the identified vulnerabilities have to be developed, to identify relevant, effective and efficient mitigations for enhancing the resilience and robustness of supply chains, particularly in critical sectors. Counterfactual model-based simulations provide useful insights in the trade-off mechanics. Finally, a continuous real-time uncertainty assessment and monitoring, collating data and intelligence across allies and partners will be utmost important in the face of the rapidly growing and dynamically changing hybrid threats. This can be done, for example, by deploying a private blockchain such as Hyperledger to establish a securely shared oversight of GSC transactions in the most critical sectors, allowing the Alliance to respond quickly when new risks and ambiguities – such as the energy weaponisation against Europe by Russia – emerge.

Among limitations, the presented analysis has not addressed the fundamental trade-off of shorter supply chains, which have lower marginal cost but also fewer suppliers, which can reduce competition, product availability, and consumer access to products. Further, under extreme parameter settings, there may be a plethora of rational expectation equilibria trajectories, without any smooth convergence properties, neither converging to a steady state or even to a limit cycle. These are promising avenues for a future research.

**Annex 1: Supply chain model**[12]

**A.1 Conceptual framework**

The world economy consists of $J \geq 1$ countries in which consumers derive utility from consuming differentiated varieties of a single sector good. Preferences across varieties are CES. The country-$i$ composite, which is used both for the final consumption, as well as to provide inputs to other firms, is a CES aggregate over the set of varieties on the unit interval:

$$Q_i = \left( \int q_i(\omega)^{(\sigma-1)/\sigma} d\omega \right)^{\sigma/(\sigma-1)}$$

(1)

where $q_i(\omega)$ denotes the quantity of variety $\omega$ that is ultimately purchased in country $i$, sourced from the lowest-cost source countries. Note that the same CES aggregator over varieties applies to the composite good, whether it is being consumed in final demand or being used as an intermediate input.

The model captures important aspects of GSCs. On the sourcing side, countries not only import consumer goods, but also intermediate inputs from various industries and countries, with these imported inputs embodying foreign value added. On the foreign market side, countries not only export consumer goods, but also intermediate inputs, thus generating domestic value added in production and exports of foreign countries.

As regards the production technology, the final demand good is produced by combining two production stages that need to be performed sequentially. Production in the initial stage $n = 1$ only uses labour, while the second stage of production combines labour with the intermediate input good produced in the first stage. More specifically, GSC production technologies can be expressed as follows:

$$y_i^1(\omega) = z_i^1(\omega) l_i^1(\omega)$$

(2)

$$y_i^2(\omega) = (z_i^2(\omega) l_i^2(\omega))^{\alpha_2} (y_i^1(\omega))^{1-\alpha_2}$$

(3)

where $\alpha_2 \in (0,1)$ denotes the labour share in stage-2 production, and $z_i^n(\omega)$ is labour productivity at stage $n$ in country $i$. Firms are perfectly competitive and the optimal location $l(n) \in \{1,\ldots,J\}$ of the different stages $n \in \{1,2\}$ of the supply chain is determined by a cost minimisation.

Countries differ in three key aspects: (i) their size, as reflected by the measure $L_i$ of 'equipped' labour available for production in each country $i$; labour is inelastically supplied and commands wage $w_i$, (ii) their geography, as captured by a $J \times J$ matrix of iceberg trade cost $\tau_{ij} \geq 1$, and (iii) their technological efficiency, as determined by the labour productivity terms $z_i^n(\omega)$. Following Eaton and Kortum (2002), we assume that $z_i^n(\omega)$ is drawn independently (across stages) from a Fréchet distribution with a cumulative distribution function $F_i^n(z) = \exp\{-T_i^n z^{-\theta}\}$.[13]

---

[12] The model based on Antràs and de Gortari (2020).

[13] The GSC complexity and opaqueness, and incomplete information regarding upstream suppliers discussed in section 2 provide a justification for the Fréchet distribution.

Consider the final demand good firm's problem in a vertically integrated supply chain of choosing the least-cost path of production to deliver consumption good variety $\omega$ to consumers in country $j$. Given equations (2) and (3), this amounts to choosing locations $l_j^1$ and $l_j^2$ to minimise

$$c(l_j^1, l_j^2) = \left(\frac{w_{l_j^2}}{z_i^2(\omega)}\right)^{\alpha_2} \left(\frac{\tau_{l_j^1}\tau_{l_j^2}w_{l_j^1}}{z_i^1(\omega)}\right)^{1-\alpha_2}$$

(4)

Given the Fréchet distribution on the labour productivity $z_i^1(\omega)$ and $z_i^2(\omega)$, yields a distribution for the equilibrium marginal cost of production of the GSCs, and facilitates a description of the general equilibrium solution. Note that due to non-linearities the minimum cost (4) associated with a given GSC path cannot be characterised by an analytically tractable solution. To gain tractability and allow the characterisation of some of the equilibrium features, we follow Antràs and de Gortari (2020) and treat the overall (i.e., chain-level) unit cost of production of a GSC flowing through a sequence of countries as a draw from a Fréchet random variable with a location parameter that is a function of the states of technology and wage levels of all countries involved in the GSC, as well as of the trade costs incurred in that supply chain.[14] Consider a given production path $\boldsymbol{l} = \{l_j^1, l_j^2\} \in J^2$, where $J$ denotes the set of countries in the world. The given supply chain's production cost is a function of trade costs, composite factor costs and the state of technology of the various countries involved in the chain. In a real world, however, two supply chains flowing across the same countries in the exact same order may not achieve the same overall productivity due to idiosyncratic factors, such as compatibility problems, production delays, or simple mistakes.

Following Antràs and de Gortari (2020) and given the cost function given in (4), we assume that the overall productivity of a given chain $\boldsymbol{l} = \{l_j^1, l_j^2\}$ is characterised by

$$\Pr((z_i^1(\omega))^{1-\alpha_2}(z_i^2(\omega))^{\alpha_2} \leq z) = \exp\left\{-z^{-\theta}\left(T_{l^1}^1\right)^{1-\alpha_2}\left(T_{l^2}^2\right)^{\alpha_2}\right\}$$

(6)

which is equivalent to assuming that $(z_i^1(\omega))^{1-\alpha_2}(z_i^2(\omega))^{\alpha_2}$ is distributed Fréchet with a shape parameter $\theta$, and a location parameter that is a function of the states of technology in all countries in the chain, as captured by $\left(T_{l^1}^1\right)^{1-\alpha_2}\left(T_{l^2}^2\right)^{\alpha_2}$. A direct implication of this assumption is that the unit cost associated with serving consumers in a given country $j$ via a given supply chain $\boldsymbol{l}$ is also distributed Fréchet, which then allows to characterise equilibrium prices and the relative prevalence of different GSCs.

The share of country $j$'s income spent on the final demand good produced under a particular GSC production path $\boldsymbol{l} \in J^2$ is given by

$$\pi_{lj} = \frac{\left(\left(T_{l^1}^1\right)^{\alpha_n}((w_{l^1})^{\alpha_n}\tau_{l^1l^2})^{-\theta}\right)^{1-\alpha_2} \times \left(T_{l^2}^2\right)^{\alpha_2}\left((w_{l^2})^{\alpha_2}\tau_{l^2j}\right)^{-\theta}}{\sum_{l \in J^2}\left(\left(T_{l^1}^1\right)^{\alpha_n}((w_{l^1})^{\alpha_n}\tau_{l^1l^2})^{-\theta}\right)^{1-\alpha_2} \times \left(T_{l^2}^2\right)^{\alpha_2}\left((w_{l^2})^{\alpha_2}\tau_{l^2j}\right)^{-\theta}}$$

(7)

and the exact ideal price index $P_j$ in country $j$ is given by

$$P_j = \kappa \left( \sum_{l \in J^2} \left( (T_{l^1}^1)^{\alpha_n} ((w_{l^1})^{\alpha_n} \tau_{l^1 l^2})^{-\theta} \right)^{1-\alpha_2} \times (T_{l^2}^2)^{\alpha_2} ((w_{l^2})^{\alpha_2} \tau_{l^2 j})^{-\theta} \right)^{-1/\theta}$$

(8)

where $\kappa$ is a constant that depends only on $\sigma$ and $\theta$. Note that for the price index to be well defined, we need to impose $\sigma - 1 < \theta$.

---

[14] The Fréchet random variable accounts for idiosyncrasies characterising real-world supply chains. For discussion, see section 2.

Equations (7) and (8) have a number of implications. On the one hand, GSCs that involve countries with higher states of technology, $T_i$, or lower labour costs, $w_i$, will tend to participate disproportionately in production paths leading to consumption in $j$. On the other hand, high trade costs penalise the participation of countries in GSCs. To see the compounding effect of trade costs, presume all trade costs in a particular GSC increase by 10%, this GSC's spending share decreases by $\theta(2 - \alpha_2)$ percent. Another implication of this compounding effect is that, in choosing their optimal path of production, firms will be more concerned about reducing trade costs in relatively downstream stages than in relatively upstream stages, as reflected in the higher exponent for $\tau_{l^1 j}$ than for $\tau_{l^1 l^2}$ in equation (7).

Next, we solve for equilibrium wages. Note that for all GVCs, stage-$n$ value added (labour income in our model) accounts for a share $\alpha_n \beta_n$ of the value of the finished good emanating from that GVC. Furthermore, the total spending in any country $j$ is given by $w_j L_j$, and the share of that spending by country j flowing to GVCs in which country $i$ is in position $n$ is given by $Pr(\Lambda_i^n, j) = \sum_{l \in \Lambda_i^n} \pi_{lj}$ where $\Lambda_i^n = \{l \in J^N | l^n = i\}$ and $\pi_{lj}$ is given in equation (7). It thus follows that the equilibrium wage vector in country $i$ is determined by the solution of the following system of equations

$$w_i L_i = \sum_{j \in J} \sum_{n \in N} \alpha_n \beta_n \times Pr(\Lambda_i^n, j) \times w_j L_j$$

(9)

The system of equations is nonlinear because $Pr(\Lambda_i^n, j)$ is a nonlinear function of wages themselves, and of the vector $\boldsymbol{P}$, which is in turn a function of the vector of wages $\boldsymbol{w}$. In a GSC with only one production stage, $N = 1$, equation (9) implies that $\alpha_n \beta_n = 1$ and $Pr(\Lambda_i^n, j) = \pi_{ij} = (\tau_{ij} c_i)^{-\theta} T_i^1 / \sum_k (\tau_{kj} c_k k)^{-\theta} T_k^1$.

## A.2 Empirical implementation and gains from participation in GSC

Although the 'GSC trade shares' in (7) are not directly observable in trade data, one can derive from them closed-form expressions for the various matrices and vectors (e.g., final-good and intermediate input flows across countries) of an aggregate (total flows aggregated across sectors) World Input Output Tables (WIOT).

First, consider final demand flows, $F$, (see Figure A.1). According to our model, for final demand goods to flow from source country $i$ to destination country $j$, it must be the case that country $i$ is in position $N$ in a supply chain serving consumers in country $j$. Defining the set of GSCs flowing through country $i$ at position $n$ by $\Lambda_i^n \in J^{N-1}$, the overall share of spending in country $j$ on goods value added in country $i$ (i.e., in GSCs in which country $i$ produces stage $N$) can be expressed as

$$\pi_{ij}^F = \frac{\sum_{l\in\Lambda_i^N} \prod_{n=1}^{N-1}\left(\left(T_{l^n}^n\right)^{\alpha_n}\left((c_{l^n})^{\alpha_n}\tau_{l^n l^{n+1}}\right)^{-\theta}\right)^{\beta_n} \times \left(T_i^N\right)^{\alpha_N}\left((c_i)^{\alpha_N}\tau_{ij}\right)^{-\theta}}{\sum_{l\in J^N} \prod_{n=1}^{N-1}\left(\left(T_{l^n}^n\right)^{\alpha_n}\left((c_{l^n})^{\alpha_n}\tau_{l^n l^{n+1}}\right)^{-\theta}\right)^{\beta_n} \times \left(T_{l^N}^N\right)^{\alpha_N}\left((c_{l^N})^{\alpha_N}\tau_{l^N j}\right)^{-\theta}}$$

(10)

where the composite cost, $c_i$, in country $i$ is captured by a Cobb-Douglas aggregator, $c_i = (w_i)^\gamma (P_i)^{1-\gamma}$ with $P_i$ being the ideal price index associated with preferences. Equation (10) implies that final demand trade flows between any two countries $i$ and $j$ are then simply given by $\pi_{ij}^F \times w_j L_j$. Although trade imbalances are not considered in this version of the model, they can be straightforwardly introduced (see e.g. Antràs and de Gortari 2020).

Analogously, based on the 'GSC trade shares' in (7) closed-form expressions for intermediate-input trade flows between any two countries $i$ and $j$ can be derived

$$\pi_{ij}^T = \frac{\prod_{n=1}^{N-1}\left(\left(T_{l^n}^n\right)^{\alpha_n}\left((c_{l^n})^{\alpha_n}\tau_{l^n l^{n+1}}\right)^{-\theta}\right)^{\beta_n} \times \left(T_{l^N}^N\right)^{\alpha_N}\left((c_{l^N})^{\alpha_N}\tau_{l^N j}\right)^{-\theta}}{\sum_{l\in J^N} \prod_{n=1}^{N-1}\left(\left(T_{l^n}^n\right)^{\alpha_n}\left((c_{l^n})^{\alpha_n}\tau_{l^n l^{n+1}}\right)^{-\theta}\right)^{\beta_n} \times \left(T_{l^N}^N\right)^{\alpha_N}\left((c_{l^N})^{\alpha_N}\tau_{l^N j}\right)^{-\theta}}$$

(11)

Note that to implement empirically, equation (11) is more involved than equation (10), as one needs to take into account both vertical trade between two contiguous stages as well as intermediate input trade flows associated with the use of the bundle of inputs at each stage (intermediate use, $T$, in Figure A.1).

Using expressions (10) and (11), we are able to estimate the key parameters of the model via maximum likelihood by minimising the distance between various moments of the WIOT and their model counterparts.

| | | Intermediate use ($T$) | | | | Final demand ($F$) | | Gross output ($X$) |
|---|---|---|---|---|---|---|---|---|
| | | Nation A | | Nation B | | Nation A | Nation B | |
| | | Sector 1 | Sector 2 | Sector 1 | Sector 2 | | | |
| Nation A | Sector 1 | $T_{1A1A}$ | $T_{1A2A}$ | $T_{1A1B}$ | $T_{1A2B}$ | $F_{1AA}$ | $F_{1AB}$ | $X_{1A}$ |
| | Sector 2 | $T_{2A1A}$ | $T_{2A2A}$ | $T_{2A1B}$ | $T_{2A2B}$ | $F_{2AA}$ | $F_{2AB}$ | $X_{2A}$ |
| Nation B | Sector 1 | $T_{1B1A}$ | $T_{1B2A}$ | $T_{1B1B}$ | $T_{1B2B}$ | $F_{1BA}$ | $F_{1BB}$ | $X_{1B}$ |
| | Sector 2 | $T_{2B1A}$ | $T_{2B2A}$ | $T_{2B1B}$ | $T_{2B2B}$ | $F_{2BA}$ | $F_{2BB}$ | $X_{2B}$ |
| Value Added ($V$) | | $V_{1A}$ | $V_{2A}$ | $V_{1B}$ | $V_{2B}$ | | | |
| Gross output ($X$) | | $X_{1A}$ | $X_{2A}$ | $X_{1B}$ | $X_{2B}$ | | | |

**Figure A.1: A schematic World Input Output Table**

The outlined general equilibrium framework allows to study the real income implications of participation in GSCs and international trade. To gain intuition, it is useful to consider a 'purely-domestic' supply chain that performs all stages in a given country $j$ to serve consumers solely in the same country $j$. Further, assume that there are N sequential production stages. Let us denote this domestic chain by $j = (j, j, \ldots, j)$. Plugging the expression for the price index in (8) into equation (7) and simplifying terms yields

$$\frac{w_j}{P_j} = \left(\kappa(\tau_{jj})^{\sum_{n=1}^{N} \beta_n}\right)^{-1} \left(\frac{\prod_{n=1}^{N}(T_j^n)^{\alpha_n \beta_n}}{\pi_{jj}}\right)^{1/\theta}$$

(12)

The gains from trade expression (12) is analogous to the Eaton and Kortum (2002) framework with one notable difference that $\pi_{jj}$ is not the aggregate share of spending on domestic intermediate or final goods (which are readily available in input output datasets), but rather the share of spending on goods that are produced entirely through domestic supply chains. A direct implication of this feature of the model is that the standard sufficient statistic approach is not applicable, and one needs to estimate the model structurally to back out $\pi_{jj}$ from the world input output data.